%% file: main.tex
\documentclass[conference]{IEEEtran}

\pdfoutput=1

\newlength{\figwidth}
\usepackage{color}
\definecolor{links}{rgb}{0.7,0,0}   
\definecolor{urls}{rgb}{0,0,0.8}    
\definecolor{cites}{rgb}{0,0,0.8}   

\usepackage[colorlinks,hyperindex,linkcolor=links,citecolor=cites,urlcolor=urls]{hyperref} 
 
\usepackage[nosort]{cite}        
\usepackage{url} 
\usepackage[intlimits]{amsmath}
\usepackage{bbm}
\usepackage{graphicx}
\usepackage{paralist}
\usepackage{fancyref}
\usepackage[stretch=16,shrink=16,step=4]{microtype}
\usepackage{vmr-symbols-vecbold}
\usepackage{standard-macros}

\usepackage{siunitx}
\makeatletter
\def\@IEEEinterspaceratioM{0.265}
\def\@IEEEinterspaceMINratioM{0.1651}
\def\@IEEEinterspaceMAXratioM{0.38}

\def\@IEEEinterspaceratioB{0.31}
\def\@IEEEinterspaceMINratioB{0.19}
\def\@IEEEinterspaceMAXratioB{0.38}
\@IEEEtunefonts
\makeatother
\usepackage{glossaries}
\glsdisablehyper
\loadglsentries{glossary}

\setacronymstyle{long-short}
\usepackage[mode=buildnew]{standalone}

\safemath{\Np}{N\sub{p}}
\safemath{\qiup}{q_{i}^{\text{up}}}
\safemath{\qilow}{q_{i}^{\text{low}}}
\safemath{\qup}{\vecq^{\text{up}}}
\safemath{\qlow}{\vecq^{\text{low}}}
\safemath{\Ess}{E\sub{\text{s}}}
\safemath{\Ed}{E\sub{\text{d}}}

\renewcommand{\markred}{}
\renewcommand{\markblue}{}
\begin{document}

\IEEEoverridecommandlockouts

\title{Deep-Learning-Based Channel Estimation for Distributed MIMO
	with 1-bit Radio-Over-Fiber Fronthaul}
%
%
%

%
\author{\IEEEauthorblockN{Alireza Bordbar$^1$, Lise Aabel$^{1,2}$,  Christian H\"ager$^1$, Christian Fager$^1$, and Giuseppe Durisi$^1$}\\
	\IEEEauthorblockA{
		$^1$Chalmers University of Technology, 41296 Gothenburg, Sweden\\
		$^2$Ericsson Research, 41756 Gothenburg, Sweden\\
	}\thanks{This work was supported in part by the Swedish Foundation for Strategic
		Research (SSF), under grants ID19-0036 and FUS21-0004. The simulations were  enabled by resources provided by the National Academic Infrastructure for Supercomputing in Sweden (NAISS), partially funded by the Swedish Research Council through grant agreement no. 2022-06725.}}
%
%
\maketitle

\begin{abstract}
	We consider the problem of pilot-aided, uplink channel estimation in a distributed massive
	multiple-input multiple-output (MIMO) architecture, in which the access points are connected
	to a central processing unit via fiber-optical fronthaul links, carrying a
	two-level-quantized version of the received analog radio-frequency signal.
	We adapt to this architecture the deep-learning-based channel-estimation algorithm recently proposed by
	Nguyen \emph{et al.} (2023), and explore its robustness to the additional
	signal distortions (beyond $1$-bit quantization) introduced in the considered
	architecture by the automatic gain controllers (AGCs) and by the comparators.
	These components are used at the access points to generate the
	two-level analog waveform from the received signal.
	Via simulation results, we illustrate that the proposed channel-estimation method outperforms the Bussgang linear minimum mean-square error channel estimator, and
	it is robust against the additional impairments introduced by the AGCs and the
	comparators.

\end{abstract}
\section{Introduction}
In distributed massive \gls{mimo}, a large number of remotely located \glspl{ap} serve a much
smaller number of \glspl{ue} in a coordinated way.
Coordination is enabled by a \gls{cpu}, which is connected to the \glspl{ap} via fronthaul
links.
With such an architecture, one can exploit macro-diversity and mitigate path-loss
variations compared to co-located massive \gls{mimo} architectures.
This yields a more
uniform quality of service~\cite{demir20-a}.

In the analysis of distributed MIMO systems, it is often assumed that the \glspl{ap} are
equipped with local oscillators, which are used for up- and down-conversion, and that
samples of the \gls{bb} signals are exchanged between the \glspl{ap} and the \gls{cpu} over the fronthaul links.
However, these local oscillators need to be synchronized for reciprocity-based joint coherent
downlink beamforming to work, which may be costly or even unfeasible in certain
deployments~\cite{larsson24-01a}.

To solve this issue, an alternative architecture has been proposed in the literature (see,
e.g.,~\cite{cordeiro17-11c,wang19-06a,wu20-02a,sezgin21-02a,aabel23-12a}), in which up-
and down-conversion are performed digitally at the \gls{cpu}.
This, however, implies that \gls{rf} signals need to be exchanged over the fronthaul links.
To limit complexity and fronthaul requirements, in the architecture proposed
in~\cite{cordeiro17-11c,wang19-06a,wu20-02a,sezgin21-02a,aabel23-12a}, only a
two-level representation of the \gls{rf} signal is exchanged over the fronthaul link.
This implies that quantizers with just a single-bit resolution can be deployed at the
\gls{cpu}.
The testbed measurements reported in, e.g.,~\cite{sezgin21-02a,aabel23-12a} demonstrate
that satisfactory performance can be achieved despite the nonlinearities introduced by
this architecture, provided that the $1$-bit quantizers at the \gls{cpu} operate at a
sufficiently high sampling rate---roughly two orders of magnitude larger than the
bandwidth of the transmitted signal.

This paper focuses on the problem of acquiring accurate channel estimates
via uplink pilot transmission, to enable reciprocity-based coherent downlink
beamforming.
The problem of channel estimation in the presence of the impairments caused by $1$-bit quantization
is well-studied in the literature, although most works focus on the scenario in which sampling
is performed on the \gls{bb} signal and not the \gls{rf} signal.
Relevant contributions involve the derivation of the \gls{lmmse} estimator via Bussgang
decomposition \cite{li17-08a} as well as of methods for the efficient evaluation of the maximum \emph{a posteriori} and \gls{ml} channel estimators, by
exploiting the convexity of the log-likelihood functions~\cite{choi15-07a,studer16-06a}.
More recently, several methods relying on deep neural networks have been proposed in the literature
(see, e.g., \cite{nguyen23-01a} and references therein).

In this paper, we will focus on the machine-learning algorithm proposed
in~\cite{nguyen23-01a}, where deep unfolding of the iterations of a first-order
optimization method is used to tackle the \gls{ml} channel-estimation problem.
Specifically, we adapt the algorithm proposed in \cite{nguyen23-01a} to the 1-bit
radio-over-fiber fronthaul architecture considered in this paper.
Our extension accounts for the real-valued nature of the sampled signals,
and for oversampling and dithering.
Furthermore, we analyze the robustness of the resulting algorithm to the additional
impairments introduced in our architecture by the \gls{agc}, which limits the range of the
received signal power at which dithering is effective, and by the comparator, which
introduces random bit flips when the difference between the signals at its input ports is
small.

\begin{figure*}[t]
	\centering
	\includegraphics[width=\textwidth]{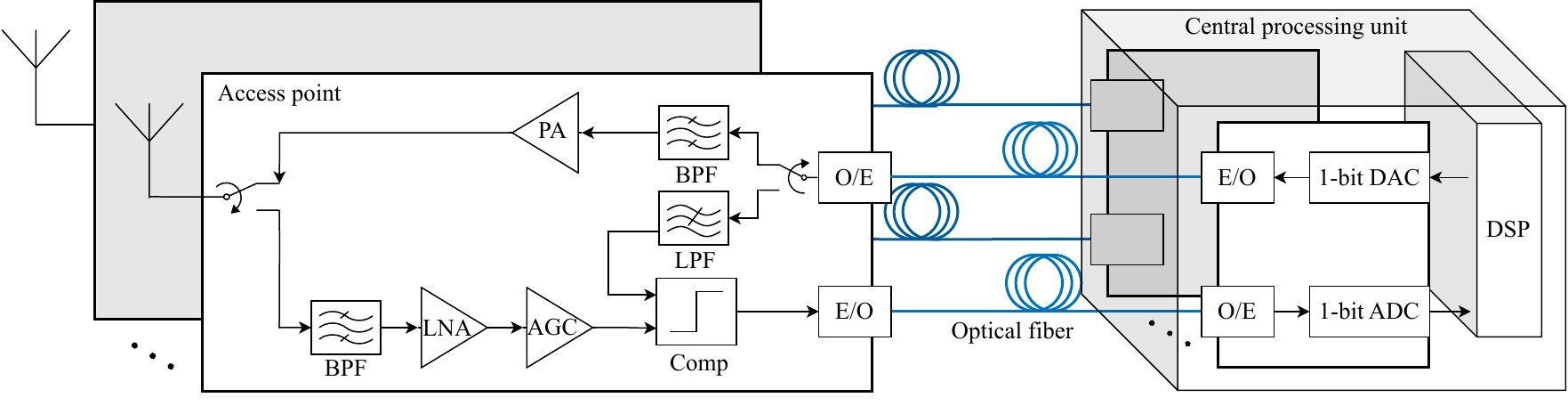}
	\caption{Block diagram of one \gls{ap} and the \gls{cpu} in the distributed \gls{mimo} architecture with 1-bit radio-over-fiber fronthaul considered in the paper.}
	\label{fig:blockdiag}
\end{figure*}
%

\section{System Model}
\subsection{Distributed MIMO with 1-Bit Radio-Over-Fiber Fronthaul}
In Fig.~\ref{fig:blockdiag}, we illustrate the key components of the distributed
\gls{mimo} architecture with 1-bit radio-over-fiber fronthaul, and single-antenna \glspl{ap}, described
in~\cite{sezgin21-02a,aabel23-12a}.
\paragraph*{Downlink}
In the downlink, modulated and precoded \gls{bb} signals intended for the \glspl{ue} are
up-converted digitally to the carrier frequency at the \gls{cpu}, and then \gls{bp}
sigma-delta modulated.
This yields an oversampled binary representation of the \gls{rf} signals.
Sigma-delta modulation relies on oversampling and noise shaping to place the quantization
distortion outside the bandwidth of the desired signal~\cite{pavan17}.
Analog signals are then generated via 1-bit \gls{dacs}.
Subsequently, these signals are converted into the optical domain by \gls{eo} converters and
conveyed to the desired \glspl{ap} via optical fibers.
A reconstruction \gls{bpf} at the \gls{ap} suppresses the out-of-band distortion produced
by the sigma-delta modulator and recovers the underlying (unquantized) \gls{rf} waveform,
which is then transmitted over the single antenna of each \gls{ap}.
\paragraph*{Uplink}
The signal received at the antenna port of each \gls{ap} is fed to a \gls{bpf} and then
passed through a \gls{lna} and an \gls{agc}.
The output of the \gls{agc} is fed to the positive differential input port of a comparator
(denoted in the figure by ``Comp'').
The negative differential input port of the comparator is fed with a dither signal, which
is provided to the \gls{ap} via the downlink fiber-optical fronthaul.
Specifically, this dither signal is generated digitally at the \gls{cpu}, sigma-delta
modulated, and recovered at the \gls{ap} via a \gls{lpf}. The \gls{agc} has the important
function to regulate the power of the received signal, in order to make dithering
effective.
The output of the comparator is converted into the optical domain and then conveyed to the
\gls{cpu}, where it is converted back to the electrical domain and then digitalized via a
$1$-bit \gls{adc}.
The resulting samples are then digitally down-converted and further processed at the
\gls{dsp} unit.

The testbed described in~\cite{aabel23-12a} has a maximum bandwidth of \SI{100}{MHz},
operates at a carrier frequency of \SI{2.35}{GHz}, and involves a sampling rate of
\SI{10}{GS/s}.
As detailed in~\cite{aabel23-12a}, this architecture can be implemented using low-cost, off-the-shelf components.

\subsection{Pilot-Aided Channel Estimation: a Mathematical Model}\label{sec:model}
We next consider the problem of uplink pilot-based channel estimation in the distributed \gls{mimo} architecture described in Fig.~\ref{fig:blockdiag}, and provide a
mathematical model for the received signal, which we will use in Section~\ref{sec:algo} to generalize the channel-estimation algorithm proposed in~\cite{nguyen23-01a}.

To keep the notation compact, we will focus on the problem of estimating the channel
between one arbitrary \gls{ap} and $U$ \glspl{ue}.\footnote{The mathematical model we
	provide in this section can be readily extended to the joint estimation of the channel between an
	arbitrary number of \glspl{ap} and \glspl{ue}. However, such generalization appears to
	be superfluous, since, in distributed \gls{mimo}, the
	channels corresponding to different \glspl{ap} are typically assumed to be independent.}

We assume that the \glspl{ue} transmit pilot signals of bandwidth~$W$, centered at a frequency
$f\sub{c}\gg W$.
We let $f\sub{s}\geq 2f_{c}+W$ denote the sampling rate at the \glspl{adc} and also assume that pilot transmission involves $\Np$ signals (where, as we shall see, $\Np$ is related
to the number of pilot symbols transmitted per \gls{ue}) of duration $T=N/f\sub{s}$ seconds, for
some integer $N$.
Under these assumptions, we can write the resulting $N\times \Np$ samples at the output of the
$1$-bit \gls{adc} as
\begin{equation}\label{eq:quantized-output}
	\matZ = \mathrm{sgn}\lefto(\matY\supp{rf} + \matD\right).
\end{equation}
Here, the $N\times \Np$ matrix $\matY\supp{rf}$ contains the samples of the discrete-time
\gls{rf} signal received at the \gls{ap} (after the \gls{bpf}), taken at the sampling frequency $f\sub{s}$;
the $N\times \Np$ matrix $\matD$ contains the samples of the dither signal added at the
\gls{ap}; finally, $\mathrm{sgn}$ denotes the sign function, which models $1$-bit
quantization, and is applied entrywise to the matrix $\matY\supp{rf}+\matD$.
For analytical convenience, we shall model the entries of the dither-signal matrix as
independent $\normal(0, E\sub{d}/2)$ random variables.
Furthermore, we model~$\matY\supp{rf}$ as
\begin{equation}\label{eq:received-rf-no-quant}
	\matY\supp{rf} = \sqrt{2}\Re\lefto\{ \diag(\vecu)\matY\supp{bb} \right\}
\end{equation}
where the $N$-dimensional vector $\vecu$, whose $n$th entry is given by $u_{n}=e^{j2\pi
	(f\sub{c}/f\sub{s})n}$, $n=0,\dots, N-1$, models the up-conversion operation, and $\matY\supp{bb}$ is
the complex envelope of the discrete-time received signal.

We let $S=WT$, and assume for simplicity that $S$ is an odd integer.
Furthermore, we define the following set:
\begin{multline}\label{eq:setS}
	\setS = \{0,1,\dots, (S-1)/2, N-(S-1)/2, N-(S-1)/2+1,\\ \dots, N-1 \}.
\end{multline}
To account for oversampling, we model $\matY\supp{bb}$ as
\begin{equation}\label{eq:complex-envelope}
	\matY\supp{bb} = \matF\supp{inv} (\matH \matP + \matW).
\end{equation}
Here, the $S\times U$ matrix $\matH$ contains the channel coefficients (expressed in the
frequency domain) that need to be estimated; the $U\times \Np$ matrix \matP contains the
pilot symbols, which we assume have power $E\sub{s}$; the $S \times \Np $ matrix $\matW$, whose entries
are drawn independently from a $\jpg(0, N_{0})$ distribution, denotes the additive
Gaussian noise; finally, the $N\times S$ matrix $\matF\supp{inv}$ is a truncated \gls{idft}
matrix, obtained by removing all columns of a $N\times N$ \gls{idft} matrix whose indices do
not belong to the set $\setS$ defined in~\eqref{eq:setS}.
Note that, in our notation, the oversampling ratio is given by $N/S$.

It will turn out convenient to vectorize the matrix $\matZ$
in~\eqref{eq:quantized-output}.
Specifically, we set $\vecz = \mathrm{vec}(\matZ) \in \reals^{N\Np \times 1}$.
Similarly, we vectorize also $\matY\supp{rf}$, $\matD$, $\matY\supp{bb}$, $\matH$, and
$\matW$, obtaining the vectors $\vecy\supp{rf}$, $\vecd$, $\vecy\supp{bb}$, $\vech$, and $\vecw$.
Using this notation, we can equivalently express~\eqref{eq:received-rf-no-quant}
and~\eqref{eq:complex-envelope} as
\begin{equation}
	\vecy\supp{rf} = \sqrt{2}\Re\left\{ \matU\vecy\supp{bb} \right\}
\end{equation}
and
\begin{equation}
	\vecy\supp{bb} = \widetilde{\matF}\supp{inv}(\widetilde{\matP}\vech +\vecw).
\end{equation}
Here, $\matU = \matI_{\Np}\otimes \diag(\vecu)$, $\widetilde{\matF}\supp{inv} = \matI_{\Np} \otimes \matF\supp{inv}$, and  $\widetilde{\matP} = \tp{\matP} \otimes \matI_{S}$, where $\otimes$ indicates the Kronecker product.
To summarize, we can compactly express the output signal in vectorized form as
\begin{equation}
	\vecz = \mathrm{sgn}\lefto( \sqrt{2}\Re\lefto\{
	\matU \widetilde{\matF}\supp{inv}(\widetilde{\matP}\vech+\vecw)\right\} + \vecd
	\right).
	\label{eq: output}
\end{equation}
%

\section{Deep-Learning-Based
  Channel Estimation}\label{sec:algo}
\subsection{Maximum-Likelihood Channel Estimation}
We start by noting that the input--output relation~\eqref{eq: output} contains two random
vectors, i.e., $\vecw \in \opC^{S\Np \times 1}$ and $\vecd \in~\opR^{N \Np \times 1}$.
The $N \Np$ entries of $\matU \widetilde{\matF}\supp{inv} \vecw$ are correlated,
which makes calculating the \gls{ml} estimator mathematically difficult. To work around
this problem, we ignore the additive noise vector $\vecw$ while deriving the ML solution
to the channel estimation problem.\footnote{The impact of this assumption will be discussed in
	Section~\ref{sec:simulation}.}
Let
\begin{equation}
	\matM = \matU \widetilde{\matF}\supp{inv} \widetilde{\matP} = \left[\vecm_1, \ldots, \vecm_{N \Np}\right]^T.
\end{equation}
Then, by ignoring $\vecw$, we can write~\eqref{eq: output} as
\begin{IEEEeqnarray}{rCl}
	\vecz = \mathrm{sgn}\lefto( \sqrt{2}\Re\lefto\{
	\matM\vech \right\} + \vecd
	\right).
	\label{eq: system model with no additive noise}
\end{IEEEeqnarray}
Let $p(\vecz | \vech)$ be the conditional probability of $\vecz$ in~\eqref{eq: system model with no additive noise} given $\vech$.
Then the \gls{ml} channel estimate $\hat{\vech}$ is given by
\begin{IEEEeqnarray}{rCl}
	\hat{\vech} &=&   \arg \underset{\vech}{\max}  \: \:p(\vecz | \vech) \nonumber\\
	&=&  \arg \underset{\vech}{\min} \: \:  \sum_{i=1} ^{N \Np} -\text{log} \Bigg[ \Phi\Big( \sqrt{\rho}\left(\qiup - \Re \lefto\{\vecm_i^{T} \vech\right\}\right)\Big)
		\nonumber \\ &&- \> \Phi \Big(\sqrt{\rho}(\qilow - \Re\lefto\{\vecm_i^{T} \vech\right\})\Big) \Bigg].
	\label{eq: maximum likelihood problem}
\end{IEEEeqnarray}
Here, $\Phi(\cdot)$ denotes the cumulative distribution function of the normal distribution, $\rho~=~\sqrt{2 / \Ed}$, and $\qiup,\qilow \in \{-\infty,0,\infty\}$ denote the upper and lower thresholds of the quantization bin to which the $i$th entry $z_i$ of the vector $\vecz$ belongs.

It is now crucial to realize that the optimization problem in~(\ref{eq: maximum likelihood
	problem}) is convex \cite{nguyen23-01a}, \cite{concavity}, \cite{studer16-06a}. Let
$f(\vech)$ be the objective function in~(\ref{eq: maximum likelihood problem}). Since
$f(\vech)$ is convex, we can solve~(\ref{eq: maximum likelihood problem}) via an iterative
gradient descent algorithm. However, the gradient of $f(\vech)$ is undefined at certain
points \cite{nguyen23-01a}. In addition, a lack of a closed-form expression for
$\Phi(\cdot)$ makes the optimization process more involved. To address these problems, we
follow~\cite{nguyen23-01a} and approximate $\Phi(\cdot)$ as
\begin{equation}
	\Phi(x) \approx \sigma (cx) = \frac{1}{1 + e^{-cx}}
	\label{eq: phi-approx}
\end{equation}
with $c = 1.702$. Using~(\ref{eq: phi-approx}), we can approximate $f(\vech)$ as
\begin{IEEEeqnarray}{rCl}
	\widetilde{f}(\vech) &=  \sum_{i=1} ^{N \Np} & -\text{log} \Bigg[ \sigma \Big( c\sqrt{\rho}(\qiup - \Re \lefto\{\vecm_i^{T} \vech\right\})\Big)
		\nonumber \\ &&- \>  \sigma \Big(c\sqrt{\rho}(\qilow - \Re\lefto\{\vecm_i^{T} \vech\right\})\Big) \Bigg].
	\label{eq: maximum likelihood problem approximate}
\end{IEEEeqnarray}
To perform iterative gradient descent on $\widetilde{f} (\vech)$, we need to compute the vector-valued Wirtinger derivative~\cite[Cor. 5.0.1]{complexdifferential}
\begin{IEEEeqnarray}{rCl}
	\frac{\partial \widetilde{f}(\vech)}{\partial \vech^*} & = & \sum_{i = 1} ^{N \Np} c\sqrt{\rho} \Big[1 - \sigma \left(c \sqrt{\rho}(\qiup - \Re\{\vecm_i ^T\vech\})\right) \nonumber\\
		&&-
		\sigma \left(c \sqrt{\rho}(\qilow - \Re\{\vecm_i ^T\vech\})\right)
		\Big] \frac{\partial \Re \{\vecm_i^T \vech\}}{\partial \vech^*} . \IEEEeqnarraynumspace
	\label{eq: gradient of the approximate function- sum}
\end{IEEEeqnarray}
Note that
\begin{IEEEeqnarray} {rCl}
	\frac{\partial \Re \{\vecm_i^T \vech\}}{\partial \vech^*}   &= & \frac{\partial}{\partial \vech ^*} \lefto( \frac{1}{2} \left[(\vecm_i^H \vech^*)^T + \vecm_i^T \vech \right]\right) \nonumber\\
	& = & \vecm_i^*/2.
	\label{eq: gradient with respect to conjugate}
\end{IEEEeqnarray}
Setting $\qup = [q_1^{\text{up}} , \ldots, q_{N\Np}^{\text{up}}]^T$,  $\qlow = [q_1^{\text{low}} , \ldots, q_{N \Np}^{\text{low}} ]^T$ and  substituting~\eqref{eq: gradient with respect to conjugate} into~\eqref{eq: gradient of the approximate function- sum}, we can write $\partial \widetilde{f}(\vech) /\partial \vech^*$ as

\begin{IEEEeqnarray}{rCl}
	\frac{\partial \widetilde{f}(\vech)}{\partial \vech^*} & = & \frac{c \sqrt{\rho}}{2} \matM^* \Bigg[\bm1 -\sigma \Big(c \sqrt{\rho}(\qup - \Re\{\matM \vech\})\Big) \nonumber
		\\
		&&- \sigma \Big(c \sqrt{\rho}(\qlow - \Re\{\matM \vech\})\Big)
		\Bigg]
	\label{eq: our setup gradient}
\end{IEEEeqnarray}
where $\sigma (\cdot)$ is applied element-wise to its vector-valued input.

We can now write the iterative gradient descent algorithm for minimizing the objective function $\widetilde{f} (\vech)$ as
\begin{IEEEeqnarray}{rCl}
	\vech^{(\ell)} = \vech^{(\ell-1)} + \alpha ^{(\ell)} \frac{\partial \widetilde{f}(\vech)}{\partial \vech^{*}}\Big\vert_{\vech = \vech ^{(\ell -1)}}\label{eq: gradient descent}
\end{IEEEeqnarray}
where $\ell$ is the current iteration index and $\alpha^{(\ell)}$ is the step size.

\subsection{Structure of the \gls{dnn}}
Following~\cite{nguyen23-01a}, we use the deep unfolding technique \cite{Deepunfolding} to implement each iteration of (\ref{eq: gradient descent}) as a \gls{dnn} layer. Each layer $\ell\in \{ 1, \ldots , L\}$ of the network receives $\vech ^{(\ell-1)} \in \opC ^{SU \times 1}$ and calculates the gradient in~\eqref{eq: gradient descent} as
\begin{multline}
	\alpha^{(\ell)} \matM^* \Bigg[\bm1
		-\sigma \left(\beta^{(\ell)}(\qup - \Re\{\matM \vech^{(\ell-1)}\})\right) \\- \sigma \left(\beta^{(\ell)}(\qlow - \Re\{\matM \vech^{(\ell-1)}\})\right)
		\Bigg]
	\label{eq: our setup gradient-DNN}
\end{multline}
where $\alpha^{(\ell)} \in \opR$ and $\beta^{(\ell)} \in \opR$ are the trainable
parameters in the $\ell$th layer.
Note that the constants $(c \sqrt{\rho}) / 2$ and $c
	\sqrt{\rho}$ in (\ref{eq: our setup gradient}) are absorbed into the trainable parameters in (\ref{eq: our setup gradient-DNN}). Training $\alpha^{(\ell)}$ and
$\beta^{(\ell)}$ corresponds to learning the step size and controlling the accuracy of the
approximation in~(\ref{eq: phi-approx}).
\section{Refined System Model}\label{sec:refined}
As pointed out in~\cite[Figs.~11 and~12]{aabel23-12a}, the input/output relation in~\eqref{eq: output}
does not match measurement results, since it ignores the impact of the \gls{agc} in the
architecture presented in Fig.~\ref{fig:blockdiag}, as well as the distortion introduced
by the comparator.
In this section, we enhance the system model in~\eqref{eq: output} to include the effect
of these two components.
\paragraph*{\gls{agc}}
The \gls{agc} is implemented in~\cite{aabel23-12a} via a feedback-controlled variable gain
amplifier, which maintains the output power at~\SI{-53}{dBW} within a~\SI{45}{dB}
dynamic range. Specifically, the maximum gain of the \gls{agc} is \SI{15}{dB}.
Let $P\supp{rf}=\Tr\lefto(\Ex{}{\vecy\supp{rf}\tp{(\vecy\supp{rf})}}\right)/N$ be the average power of $\vecy\supp{rf}$.
We model the variable gain in \SI{}{dB} introduced by the \gls{agc} as
\begin{equation}\label{eq: agc}
	G\lefto(P\supp{rf}\right) =
	\begin{cases}
		\SI{15}{dB},                & \text{if } P\supp{rf} < \SI{-68}{dBW} \\
		\SI{-30}{dB},               & \text{if } P\supp{rf} > \SI{-23}{dBW} \\
		- P\supp{rf} - \SI{53}{dB}, & \text{otherwise}.
	\end{cases}
\end{equation}
The input to the comparator can then be modeled as $a\lefto(P\supp{rf}\right)\vecy\supp{rf} + \vecd$, where $a\lefto(P\supp{rf}\right) = 10^{G\lefto(P\supp{rf}\right)/20}$.

\paragraph*{Comparator}
As discussed in~\cite{aabel23-12a}, the comparator in Fig.~\ref{fig:blockdiag} produces a lowpass-filtered version of a two-level waveform, because of its bandwidth
limitations.
As illustrated in~\cite[Fig.~12]{aabel23-12a}, when the synthesized waveform deviates
significantly from the expected two-level waveform, the bits at the output of
the~\glspl{adc} are essentially generated at random, due to limitations in the decision
circuitry.
Typically, this occurs whenever the magnitude of the signal at the input of the comparator is
small.
To model this effect, we let $\matB$ be a diagonal matrix whose diagonal entries are drawn
uniformly at random from the set $\{-1,1\}$ if the corresponding entry of the vector $a\lefto(P\supp{rf}\right)\vecy\supp{rf}+\vecd$
is smaller in absolute value than a threshold $t$, and are set to $1$
otherwise.
In our simulations, we set $t= 2.6\cdot 10^{-4}$, because this
value matches the measurements reported in~\cite{aabel23-12a}.

To summarize, in the proposed refined system model, the input--output relation~\eqref{eq: output} is replaced by
\begin{equation}\label{eq:output-refined}
	\vecz = \matB\, \mathrm{sgn}\lefto( a\lefto(P\supp{rf}\right)\sqrt{2}\Re\lefto\{
	\matU \widetilde{\matF}\supp{inv}(\widetilde{\matP}\vech+\vecw)\right\}  +\vecd \right).
\end{equation}
\begin{table}
	\centering
	\caption{The three different uplink models used for generating data to train and test the DNN-based channel estimator.}
	\label{table: models}
	\begin{tabular}{p{1cm}|p{6.5cm}}
		Model & Description                                                                                       \\
		\hline
		1     & Simplified input--output relation  (see~(\ref{eq: system model with no additive noise}))
		\\
		\hline
		2
		      & Input--output relation without \gls{agc} effects and \gls{adc} bitflips  (see~(\ref{eq: output}))
		\\
		\hline
		3     & Refined input--output relation (see~(\ref{eq:output-refined}))
	\end{tabular}

\end{table}

%
\section{Simulation Results}
\label{sec:simulation}
\paragraph*{Simulation Parameters}
Throughout this section, we assume that a single \gls{ue} ($U=1$) transmits $\Np = 10$ pilot symbols with power $\Ess$ to the AP. We also assume that $W = 240$ MHz, $f_c = 2.4$ GHz, $f_s = 10$~GS/s, $S = 9$, and $N = 189$.
Finally, we let $\vech \sim \setC \setN (\bm0 , \matI_{S})$.
The \gls{dnn} used for channel estimation consists of \markred{$L = 7$} layers. Denoting
$\hat{\vech} = \vech^{(L)}$, we take as  loss function the $\ell_2$ norm of $\vech -
	\hat{\vech}$. We use the Adam optimizer \cite{adamoptimizer} with a learning rate that
starts at $0.002$ and is multiplied by  $0.95$ after every $100$ training epochs. The size
of each training batch is $1000$, and the number of training epochs is \markred{$1000$}. The
performance metric used to evaluate and compare the channel estimation performance is the
\gls{nmse} averaged over $5000$ channel realizations. The initial values for
$\alpha^{(\ell)}$ and $\beta^{(\ell)}$ are $1.0$ and $5.0$, respectively. Furthermore,
$\vech^{(1)} = \bm0$.

\begin{figure}
	\centering
	\includegraphics[scale = 0.8]{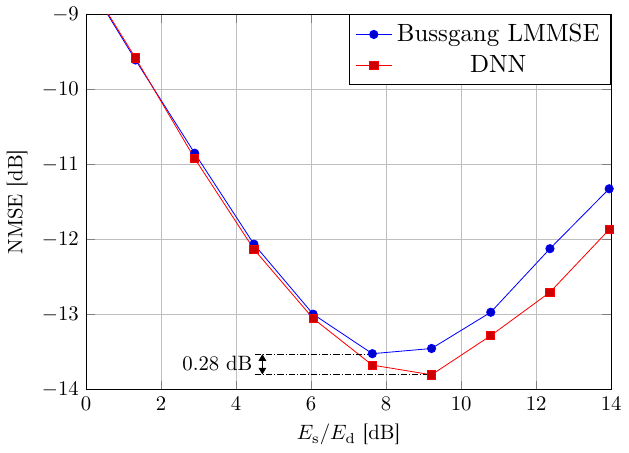}
	\caption{\gls{nmse} as a function of signal-to-dither ratio $\Ess / \Ed$ for Bussgang \gls{lmmse} and \gls{dnn}-based channel estimators.}
	\label{fig:Es_Ed}
\end{figure}

In Table~\ref{table: models}, we summarize the three different system models that are used for evaluating the performance of the proposed channel estimation algorithm.
\paragraph*{Impact of Dithering} \label{subsec : impact of dithering}
We start by considering Model~1, where we ignore the additive noise
$\vecw$, and investigate the impact of the dither signal on the \gls{nmse}. The network is
trained for $20$ $\Ess/\Ed$ values, uniformly spaced in the interval \markred{$[-5\: \text{dB} , 25\: \text{dB}]$}. The results are shown in Fig.
\ref{fig:Es_Ed}.
We see that, for Model 1,  the proposed \gls{dnn}-based channel estimator
outperforms the Bussgang \gls{lmmse} \markred{(BLMMSE)} estimator by \markred{$\qty{0.28}{dB}$ at the respective optimal $\Ess/ \Ed$ values for BLMMSE ($\qty{7.6}{dB}$) and DNN ($\qty{9.2}{dB}$)}.
\markblue{These results are in line with the ones reported in~\cite[Fig.~7(a)]{nguyen23-01a} for a different channel model.}

\paragraph*{Impact of Additive Noise}

\begin{figure}
	\centering
	\includegraphics[scale = 0.8]{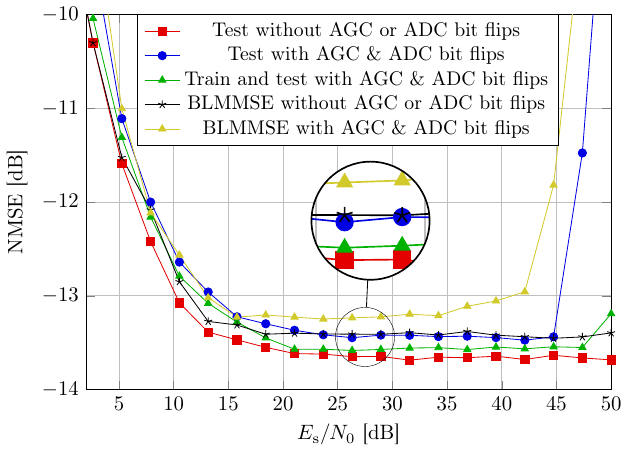}
	\caption{The effect of \gls{agc} and ADC bit flips on the performance of the proposed \gls{dnn}-based channel estimator.}
	\label{fig:SNR-NMSE-With-AGC}
\end{figure}

Next, we evaluate the test performance of a \gls{dnn} trained at the optimal $\Ess / \Ed$
value in Fig.~\ref{fig:Es_Ed} using Model~1, for the case in which the test samples are generated according to Model~2. In
Fig.~\ref{fig:SNR-NMSE-With-AGC} we show the \gls{nmse} as a function of $\Ess / N_0$. We
observe that ignoring the additive noise during the training phase of the \gls{dnn} (red curve) does
not degrade its performance when  noise is present.
Indeed, the achieved \gls{nmse} value is in agreement with the optimal $\Ess/\Ed$ value reported in
Fig.~\ref{fig:Es_Ed}
for sufficiently large \markred{$\Ess/N_{0}$}. 
\markred{The} \gls{dnn}-based \gls{ml} estimator outperforms the Bussgang
\gls{lmmse} (black curve) by around \markred{$\qty{0.22}{dB}$} \markred{on average in the interval $[\qty{20}{dB} , \qty{50}{dB}]$} when additive noise is present.
\paragraph*{Impact of \gls{agc} and Random Bit Flips}
Next, we investigate the impact of the additional impairments described in Section~\ref{sec:refined}.
We consider two additional scenarios:
\begin{inparaenum}[i)]
	\item We test the trained \gls{dnn} at the optimal $\Ess / \Ed$ obtained from
	Fig.~\ref{fig:Es_Ed}
	using Model~1 on a dataset generated from Model~3.
	\item We train the \gls{dnn} on datasets generated using Model~3, by considering $30$ $\Ess/N_{0}$ values, spaced uniformly in the interval $[\qty{0}{dB} , \qty{50}{dB}]$; the ratio $\Ess/E\sub{d}$ is still chosen according to Fig.~\ref{fig:Es_Ed}.
	Then, we test each trained \gls{dnn} on a new dataset generated  from Model~3 using the same $\Ess/N_{0}$.
\end{inparaenum}

The \gls{nmse} achieved for these two additional scenarios is also shown in Fig.~\ref{fig:SNR-NMSE-With-AGC}.
We first notice that the \gls{dnn} network trained
on the dataset that does not take into account the effects of the \gls{agc} and \gls{adc}
random bit flips (blue curve) performs poorly when $\Ess/N_0$ is \markred{outside the dynamic range of the AGC}.
Indeed, in agreement with~\eqref{eq: agc}, outside this interval, the \gls{agc} cannot
enforce the optimal signal-to-dither ratio.
Furthermore, 
\markred{averaged over the interval $[\qty{20}{dB} , \qty{40}{dB}]$}, we see \markred{around $\qty{0.22}{dB}$} performance degradation compared to the case when Model~2 is used (red curve), due to the effect of the random bit
flips.

Interestingly, we also see that if we train the \gls{dnn} on data generated according to
the refined input--output relation~\eqref{eq:output-refined} (green curve), the performance is \markblue{increased}, despite the \gls{dnn} possessing only \markred{$14$} trainable parameters,
and despite the impairments not been explicitly modeled when deriving the gradient update
step.

\section{Conclusions}
We have considered the problem of pilot-based uplink channel estimation in the distributed
\gls{mimo} $1$-bit radio-over-fiber architecture recently demonstrated in~\cite{aabel23-12a}.
Specifically, we have adapted to this architecture the deep-unfolding-based \gls{ml}
channel estimation algorithm recently proposed in~\cite{nguyen23-01a}, and analyzed its
robustness to the additional impairments introduced in the considered architecture by the
\gls{agc} (dynamic range) and the comparator (random bit flips).
In future works, 
\markblue{}
we will measure over-the-air the performance achievable with the proposed
algorithm using the testbed described in~\cite{aabel23-12a}, and explore additional
data-driven methods for channel estimation and data detection.
\bibliographystyle{IEEEtran}
\bibliography{extracted.bib}
\end{document}

%% file: glossary.tex
\newacronym{rf}{RF}{radio frequency}
\newacronym{cpu}{CPU}{central processing unit}
\newacronym{mimo}{MIMO}{multiple-input multiple-output}
\newacronym{ap}{AP}{access point}
\newacronym{dsp}{DSP}{digital signal processing}
\newacronym{bb}{BB}{base-band}
\newacronym{bpf}{BPF}{band-pass filter}
\newacronym{lpf}{LPF}{low-pass filter}
\newacronym{bp}{BP}{band-pass}
\newacronym{osr}{OSR}{oversampling rate}
\newacronym{ue}{UE}{user equipment}
\newacronym{ofdm}{OFDM}{orthogonal frequency-division multiplexing}
\newacronym{lmmse}{LMMSE}{linear minimum mean-squared error}
\newacronym{evm}{EVM}{error vector magnitude}
\newacronym{mr}{MR}{maximum ratio}
\newacronym{zf}{ZF}{zero forcing}
\newacronym{bs}{BS}{base station}
\newacronym{csi}{CSI}{channel state information}
\newacronym{dacs}{DACs}{digital-to-analog converters}
\newacronym{dac}{DAC}{digital-to-analog converter}

\newacronym{adc}{ADC}{analog-to-digital converter}
\newacronym{oe}{O/E}{optical-to-electrical}
\newacronym{eo}{E/O}{electrical-to-optical}
\newacronym{agc}{AGC}{automatic gain controller}
\newacronym{idft}{IDFT}{inverse discrete Fourier transform}
\newacronym{ml}{ML}{maximum likelihood}
\newacronym{map}{MAP}{maximum a posteriori}
\newacronym{lna}{LNA}{low noise amplifier}
\newacronym{dnn}{DNN}{deep neural network}
\newacronym{nmse}{NMSE}{normalized mean square error}
\newacronym{bmmse}{BMMSE}{Bussgang minimum mean-squared error}